\documentclass[prl,reprint,aps,superscriptaddress]{revtex4-1}

\usepackage{amsmath}	
\usepackage{amssymb}	
\usepackage{graphicx}	
\usepackage{dcolumn}	
\usepackage{bm}			
\usepackage{hyperref}
\usepackage{url}
\usepackage{lineno}

\newcommand\beq{\begin{equation}}
\newcommand\ee{\end{equation}}
\newcommand\eeq{\end{equation}}
\usepackage{color} 

\newcommand{\revise}[1]{{\color{blue}{#1}}}

\begin{document}
\title{Experimental Confirmation of the Standard Magnetorotational Instability Mechanism with a Spring-Mass Analogue}

\author{Derek M.H. Hung}
\affiliation{Department of Astrophysical Sciences, Princeton University, Princeton, New Jersey 08544, USA}
\affiliation{Princeton Plasma Physics Laboratory, Princeton, New Jersey 08543, USA}
\author{Eric G. Blackman}
\email{blackman@pas.rochester.edu}
\affiliation{Department of Physics and Astronomy, University of Rochester, Rochester, NY 14627, USA}
\affiliation{Laboratory for Laser Energetics, University of Rochester, Rochester NY, 14623, USA}
\author{Kyle J. Caspary}
\affiliation{Princeton Plasma Physics Laboratory, Princeton, New Jersey 08543, USA}
\author{Erik P Gilson}
\affiliation{Princeton Plasma Physics Laboratory, Princeton, New Jersey 08543, USA}
\author{Hantao Ji}
\email{hji@pppl.gov}
\affiliation{Department of Astrophysical Sciences, Princeton University, Princeton, New Jersey 08544, USA}
\affiliation{Princeton Plasma Physics Laboratory, Princeton, New Jersey 08543, USA}
	
\date{\today}
	
\begin{abstract}
The Magnetorotational Instability (MRI) has long been considered a plausibly ubiquitous mechanism to destabilize otherwise stable Keplerian flows to support radially outward transport of angular momentum. Such an efficient transport process would allow fast accretion in astrophysical objects such as stars and black holes to release copious kinetic energy that powers many of the most luminous sources in the universe. But the standard MRI under a purely vertical magnetic field has heretofore never been directly measured despite numerous efforts over more than a decade. Here we report an unambiguous laboratory demonstration of the spring-mass analogue to the standard MRI by comparing motion of a spring-tethered ball within different rotating flows. The experiment corroborates the theory: efficient outward angular momentum transport manifests only for cases with a weak spring in quasi-Keperian flow. Our experimental method accomplishes this in a new way, thereby connecting solid and fluid mechanics to plasma astrophysics.
\end{abstract}

\maketitle

\noindent
\textbf{\large Introduction}

Understanding angular momentum transport in astrophysical disks comprises a long standing enterprise, spanning planetary, stellar, black hole, galactic, and laboratory astrophysics. The challenge originated 250 years ago \cite{Swedenborg1734,Kant1755,Laplace1796} with enduring questions about how the angular momentum distribution within the solar system evolved from its original nebular gas \cite{Armitage2011,Kley2012,Morbidelli2016}. In addition, luminous and jetted sources in the universe, including quasars, x-ray binaries \cite{Kylafis2015,Blaes2014,Romero2017}, pre-planetary nebulae \cite{Bujarrabal2001,Blackman2001}, and gamma-ray bursts \cite{Levan2016} are likely powered by the conversion of gravitational potential energy into kinetic energy and radiation, as matter accretes onto central engines \cite{Frank2002}. Since accreting plasma typically originates far from the core of the potential well, conserving even a modest initial angular momentum during infall would prevent matter from reaching the engines. Angular momentum must be extracted much faster than microphysical diffusivities alone allow.

Enhanced transport is typically parameterized by a ``turbulent viscosity'', allowing practical accretion disk models to be compared with observations \cite{Shakura1973}. 
What mechanisms supply enhanced transport and how to model it are long standing physics problems of astrophysics \cite{Balbus2003,Blackman2015b}.
A ubiquitous source of turbulence is thought to be the magnetorotational instability (MRI) \cite{Velikhov1959,1960PNAS...46..253C} as applied to accretion discs \cite{Balbus1991,Hawley1995,Brandenburg1995,Balbus1998}: while purely hydrodynamic discs require a decreasing angular momentum gradient for linear instability, the MRI in a magnetohydrodynamic (MHD) disk requires only a radially decreasing angular velocity, so magnetized Keplerian disks of astrophysics should be unstable. Growth and saturation of the MRI are widely studied \cite{Guan2011,Flock2011,Hawley2011,Parkin2014,Bodo2014,Nauman2015,Shi2016,Bhat2016}.

The scientific method establishes scientific fact by corroborating theory with experiment, no matter how widely assumed the veracity of  a theoretically  calculated mechanism may otherwise be. As such, there are substantial efforts to demonstrate the MRI in the laboratory using differentially rotating 
liquid metals \cite{Ji2001,Sisan2004,Ji2013} and plasma \cite{Flanagan2015}, and even polymer fluids \cite{Boldyrev09,Bai2015} or an elastic beam~\cite{Vasil2015}. Purely hydrodynamic flow experiments confirm the Rayleigh criterion for stability \cite{2006Natur.444..343J,Schartman2012}. Measurements of the MRI in the standard setup with a purely vertical field in liquid metals are challenging, although recent evidence of related helical and azimuthal field MRI has been reported \cite{Stefani2009,Seilmayer2014}. The result of \cite{Sisan2004}, for example, is now understood to result from boundary effects \cite{Gissinger2012}.  There is further optimism as boundary control improves \cite{Wei2016,Caspary2018}, but so far, none of these experiments have yet demonstrated the vertical MRI.

Here we take a different approach. We appeal to the known result that the dispersion relation of the MRI for an initially vertical magnetic field also characterizes the motion of two masses tethered by a weak spring \cite{Balbus1998,Blackman2015b}. The spring represents the magnetic field and the mass represents a parcel of MHD fluid.
It has been speculated~\cite{Blackman2015b} that this analogue might be experimentally testable in the laboratory, distinct from multi-tethered configurations that have been previously theoretically explored \cite{Breakwell1981,Beletsky1985,Pizarro-Chong2008}.

Below we discuss the design and results from a new tethered ball experiment using the Princeton Taylor-Couette apparatus with water or Hydrodynamic Turbulence Experiment (HTX) \cite{Edlund14}. We compare the radial motion of the ball for cases when the ball is untethered, weakly tethered, and strongly tethered. 
As predicted by the MRI mechanism, angular momentum is transported efficiently outward only in the cases with a weak spring in quasi-Keplerian flows. The experiment demonstrates a new way to use solid and fluid mechanics to study astrophysical processes in the lab.

\begin{figure}[t]
\includegraphics[width=2.5in]{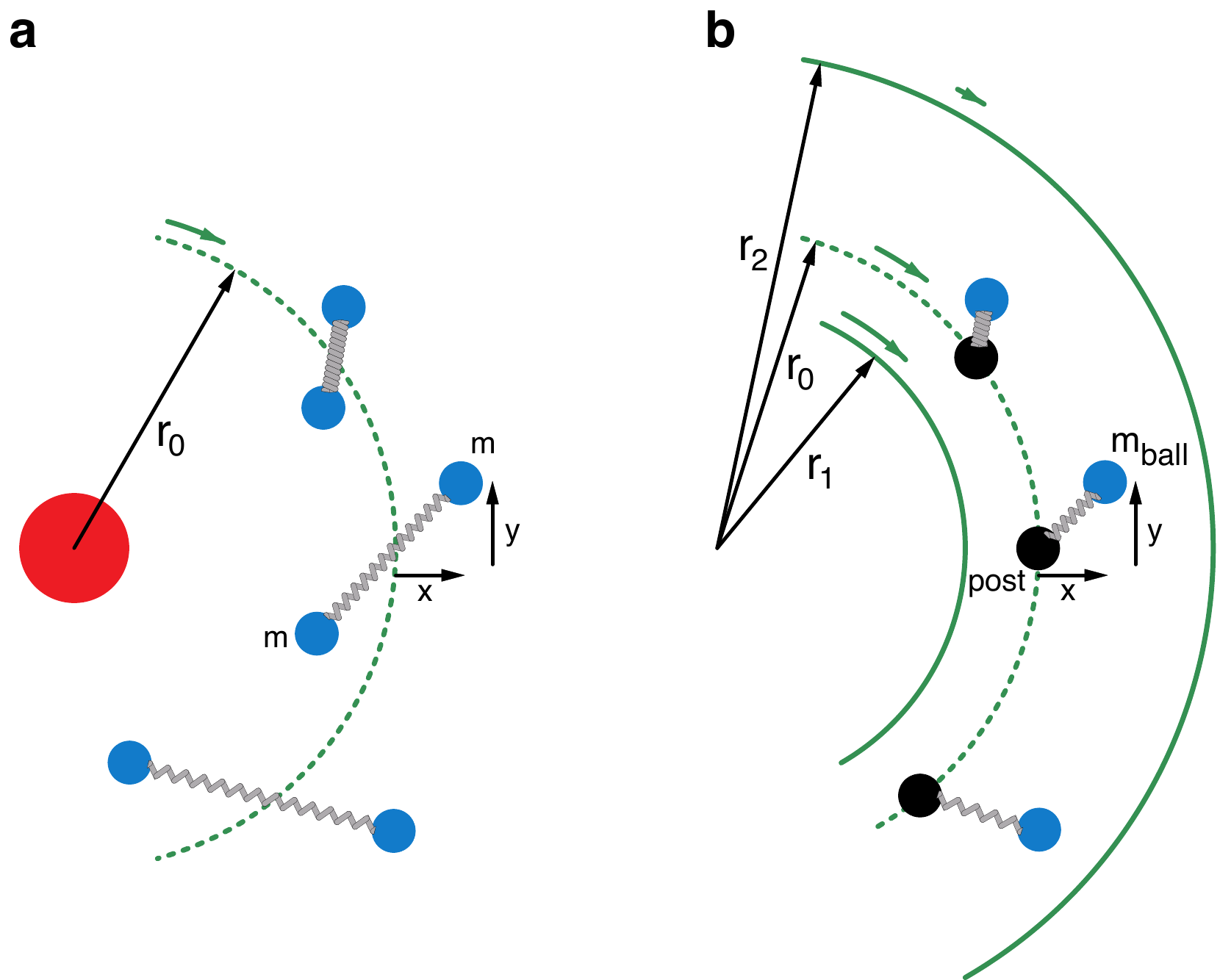}
\caption{Analogues of standard Magnetorotational Instability (MRI) in a purely vertical magnetic field. \textbf{a} Conventional MRI analogue using two equal masses (solid blue circles) tethered by a weak spring, under the influence of a central gravitational force. \textbf{b} The MRI analogue  that we study in our experiment, depicted in the lab frame.  A light mass (solid blue circle) is tethered to a fixed post (solid black circle) moving at angular speed $\mathit{\Omega}_3$ through a weak spring embedded in a Taylor-Couette flow, with inner cylinder and outer cylinder rotating at $\mathit{\revise{\Omega}}_1$ and $\mathit{\Omega}_2$, respectively.  The outward radial pressure gradient, sustained through the flow by the outer rigid wall in \textbf{b}, plays the role of the central gravity of \textbf{a}. If the inner tethered mass in \textbf{a} were much larger than the outer mass (both still being much less than the central mass) the equations for the MRI analogue depicted in the two figures would be identical.}
\label{Cartoon}
\end{figure}

\bigskip
\noindent
\textbf{\large Results}

\noindent
\textbf{Theoretical model and predictions.}  
A Keplerian flow with a weak vertical magnetic field, $B_z$, subjected to perturbations within the horizontal plane $(r, \theta)$ exhibits the MRI.
The minimalist MHD version of the equations depends only on the displacement of the field lines in the plane perpendicular to the initial magnetic field.  
Two masses tethered by a weak spring orbiting in a central potential \cite{Balbus2003} then provide an analogue of this local instability, although the minimalist MHD MRI equations
most directly correspond to the motion of a single mass tethered to a fixed point in a co-rotating frame \cite{Blackman2015b} (Fig.~\ref{Cartoon}a vs. Fig. ~\ref{Cartoon}b).

Physically, the linear phase of the instability interpreted in the context of Fig.~\ref{Cartoon}b is expected to occur as follows: a light test mass is released from a post that is fixed to orbit with the flow at angular speed $\mathit{\Omega}_3$. The mass is tethered to a weak spring.  If the spring is weak enough such that oscillation time is significantly longer than an orbit time but still strong enough to couple the post and mass over this time scale, the post will transmit angular momentum to the test mass moving the latter outward. If the spring is too strong, outward motion is limited by the spring tension, effectively retaining the ball as part of the post.

Mathematical correspondence between the minimalist MRI unstable MHD equations and those of tethered mass motion is simplest in local Cartesian coordinates $x,y,z$ in a rotating frame with radius $r=r_0+x$ and $r_0(\theta-\theta_0)=y$, with fixed point at $x=y=0$. This point moves in the lab frame with angular velocity ${\mathit{\Omega}_3}\equiv {\mathit{\Omega}}(x=0)$ and the shear flow away from the fixed point in the rotating frame is given by $r ({\mathit{\Omega}}-\mathit{\Omega}_3) \simeq x r \partial_r{\mathit{\Omega}} |_{r=r_0}= -x q \mathit{\Omega}_3$, with $q\equiv- {d \ln { \mathit{\Omega}} /d \ln r}$.  For the MHD case, when the centrifugal force is balanced by gravity and total pressure gradients are ignored, the local 2-D MHD momentum equations are
\begin{eqnarray}
{\ddot x} -2\mathit{\Omega}_3 {\dot y} & = & -(K_A- T) x,
\label{hill1a}\\
{\ddot y} +2\mathit{\Omega}_3 {\dot x} & = & -K_A y.
\label{hill2b}
\end{eqnarray}
Dots indicate time derivatives; $T= 2 q \mathit{\Omega}_3^2$ 
is the coefficient of the tidal force per unit mass; the second terms on the left sides come from the Coriolis force; $K_A=(k{v}_A)^2$, 
arises from magnetic tension where ${v}_A$ is the Alfv\'en speed associated with the vertical field. 
 
Equations~(\ref{hill1a}) and (\ref{hill2b}) also approximate motion of a mass tethered to a fixed point $x,y=0$ by a spring with spring constant per unit mass $K_A$, as in Fig.~\ref{Cartoon}(b). [For Fig.~\ref{Cartoon}(a) this requires $\mathit{\Omega}=\mathit{\Omega}_3$ and  $K_A\rightarrow 2K_A$ \cite{Blackman2015b}.]
The Coriolis and tidal force terms arise whether supplied by gravity without pressure gradients, or by pressure gradients when the mass is embedded in a laboratory quasi-Keplerian (qK) flow without gravity.
For initial displacements $[x(t) e^{ikz},y(t)e^{ikz}]$ and $q < 0$, 
the system is stable. But for $q>0$, when $K_A<T$, the MRI instability ensues.
For $K_A=0$ (no spring), the right side of Eq.~(\ref{hill2b}) vanishes and ${\dddot x}= {\dot x}(T-4\mathit{\Omega}_3^2)$. The behavior then depends on $q$: the coefficient of $\dot x$ changes sign at $q=2$, and instability occurs only for $q>2$ --- the Rayleigh unstable regime.

Although the Cartesian approximation captures the MRI mechanism, modeling the MRI mechanism with our our Taylor-Couette experiment requires inclusion of the non-linear curvature and damping terms. In cylindrical coordinates, the vector lab-frame equation of motion for a tethered mass in the rotating background flow is 
\begin{eqnarray}
\ddot{{\bf r}}
&= & {{\bf f}}_{\rm c} - {\bar K}\left[{{\bf r}(t)-{\bf r}_{\rm p}}(t)\right] \nonumber \\
& &- ({D}_1+{ D}_2 |\dot{{\bf r}}-r\mathit{\Omega}(r){\hat {\bf e}}_\theta |)\left[\dot{{\bf r}}-r\mathit{\Omega}(r){\hat {\bf e}}_\theta\right],
 \label{1}
\end{eqnarray}
where $t$ is time; ${\bf r}= r {\hat {\bf e}}_{\rm r}$ and ${\bf r}_{\rm p}= r_0 {\hat {\bf e}}_{\rm r_0}$ are the time-dependent position vectors of the ball and its launch locus (the post) respectively; $\bar K$ is the spring constant divided by the mass of the ball;  $\mathit{\Omega}(r)\simeq \mathit{\Omega}_0(r/r_0)^{-q}$, where $q$ is a constant; and
${{\bf f}}_{\rm c} = -r \mathit{\Omega}^2(r){\hat {\bf e}}_r$  is  the centripetal  force per unit mass on the ball, supplied  by the background fluid pressure gradient transmitted from the outer wall.   It is   equal and opposite in magnitude to  the  centrifugal  force per unit mass of  the flow of the local rotating frame when the background flow is in equilibrium. Quantities $D_1$ and $D_2$ are the Stokes and Reynolds drag coefficients \cite{Landau1987} given by $D_1 = 6 \pi \rho_{H_2O} \nu_{H_2O} R / M$ and $D_2 = C_D \pi \rho_{H_2O} R^2 / 2M$, for water density $\rho_{H_2O}$, kinematic viscosity $\nu_{H_2O}$, test mass radius $R$, test mass $M$, and drag coefficient $C_D$. Using $R=1.27$~cm and neutrally buoyant test mass, $D_1 = 0.0284$~s$^{-1}$ and $D_2 = 15.0$~m$^{-1}$ in our experiments.

\begin{table}[b]
\begin{center}
	\begin{tabular}{|c|c|c|c|c|c|}
	\hline
		Flow Profile & \multicolumn{2}{c|}{Solid Body} & \multicolumn{3}{c|}{Quasi-Keplerian} \\
		$(\mathit{\Omega}_1,\mathit{\Omega}_3,\mathit{\Omega}_2) [$rpm$]$ & \multicolumn{2}{c|}{(60, 60, 60)} & \multicolumn{3}{c|}{(190, 80, 22)} \\
		\hline
		Tether Strength & {none} & {weak} & {none} & {weak} & {strong} \\
		$\bar K [s^{-2}]$ & 0 & 75.4 & 0 & 75.4 & 6103.2 \\
		\hline
		4 Complex Solutions & $\pm 12.6$ & $\pm 17.0$ & $\pm 4.2$ & $\pm 15.2$ & $\pm 86.1$ \\
		$\mathit{\Omega} [$rad $s^{-1}]$ & $\pm 0.0$ & $\pm 4.4$ & $\pm 0.0$ & ${\bf \pm 7.8i}$ & $\pm 69.3$ \\
		\hline
		 \# of Experimental Runs & 4 & 4 & 8 & 8 & 4 \\
		\hline
	\end{tabular}
\end{center}
\caption{Theoretical Predictions. Four complex solutions to the linear limit of Eqs.~(\ref{hill1}) and (\ref{hill2}) [\textit{i.e.} Eqs.~(\ref{hill1a}) and (\ref{hill2b})] when variables are assumed to be proportional to $\exp (i\omega t)$ for tethered and untethered cases in solid body or quasi-Keplerian (qK) flow. The tether spring constant divided by the mass of the ball, $\bar K$, are also listed. Real values indicate oscillatory solutions, while imaginary values (boldface) indicate exponential growth and damping modes. The number of experimental runs for each case is given. The full nonlinear solutions of Eqs.~(\ref{hill1}) and (\ref{hill2}) for these cases are plotted along with experimental data in Figs.~\ref{Trajectory} and \ref{AngularMomentum}.}
\label{growth}
\end{table}

Since $d{\hat {\bf e}}_r/dt = {\dot {\theta }}{\hat {\bf e}}_\theta$, 
Eq.~(\ref{1}) contains both the azimuthal and radial components of the force equation. For initial values $r(0)=r_0; {\theta}(0)= {\theta}_{\rm p}(0)={\theta}_0;  \dot r_{\rm p} = 0,\dot {\theta}_{\rm p} = \mathit{\Omega}_3$, (where ${\theta}_{\rm p}$ is the angular coordinate of the post), the coupled equations for $r(t)$ and ${\theta}(t)$ are given by
\begin{eqnarray}
\ddot r & = & r \left[ \dot {\theta} ^2 - \mathit{\Omega}^2(r) \right] - \bar K \left[r - r_0 \cos({\theta} - {\theta}_0 - \mathit{\Omega}_3 t) \right] \nonumber \\
&& - D_1 \dot r - D_2 \left[ \dot r ^2 + r ^2 [\dot {\theta} - \mathit{\Omega}(r)] ^2 \right] ^{1/2} \dot r, \label{hill1} \\
r \ddot {\theta} & = & -2 \dot r \dot {\theta} - \bar K r_0 \sin({\theta} - {\theta}_0 - \mathit{\Omega}_3 t) - D_1 r \left[\dot {\theta} - \mathit{\Omega}(r)\right] \nonumber\\
&& - D_2 \left[ \dot r ^2 + r ^2 [\dot {\theta} - \mathit{\Omega}(r)] ^2 \right] ^{1/2} r \left[\dot {\theta} - \mathit{\Omega}(r)\right],
\label{hill2}
\end{eqnarray}
where we have used ${\hat {\bf e}}_r \cdot {\hat {\bf e}}_{r_p}=\cos({\theta}-{\theta}_0-\mathit{\Omega}_3 t)$ 
and
${\hat {\bf e}}_{\theta}\cdot {\hat {\bf e}}_{r_p}=\sin({\theta}-{\theta}_0-\mathit{\Omega}_3 t).$ 
Eqs.~(\ref{hill1}) and (\ref{hill2}) reduce to Eqs.~(\ref{hill1a}) and (\ref{hill2b}) in the linear limit.

For realistic parameters, the $D_1$ term is  small. In the linear regime,  
the $D_2$ term also does not contribute and Eqs.~(\ref{hill1}) and (\ref{hill2}) then predict runaway displacement in the usual MRI unstable regimes, namely $ 0<q<2$ and ${\bar K}>0$, but not $ 0<q <2$ and ${\bar K}=0$ (Table \ref{growth}). By choosing springs with proper strengths, the MRI mechanism can be directly tested using a tethered ball in qK flows.

We emphasize that even when $D_1$ and $D_2$ are small, the ball is still strongly coupled to the flow by the background fluid pressure forces. In the vertical direction the upward pressure force balances gravity to maintain neutral buoyancy which keeps the primary ball motion confined to 2-D. The radial pressure force transmitted from the outer wall balances the outward radial force associated with rotation as we have discussed in defining  ${\bf f}_{\rm c}$ above.

\begin{figure}[b]
\includegraphics[width=1.6in]{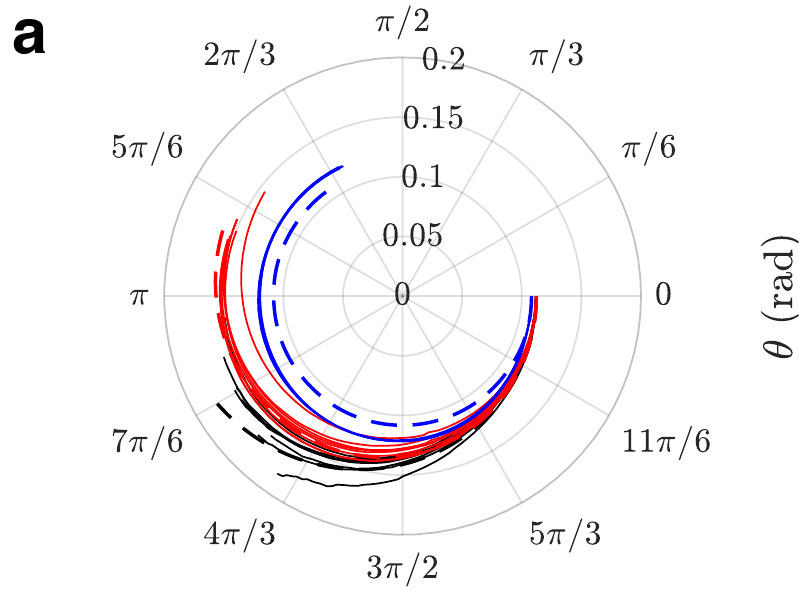}
\includegraphics[width=1.6in]{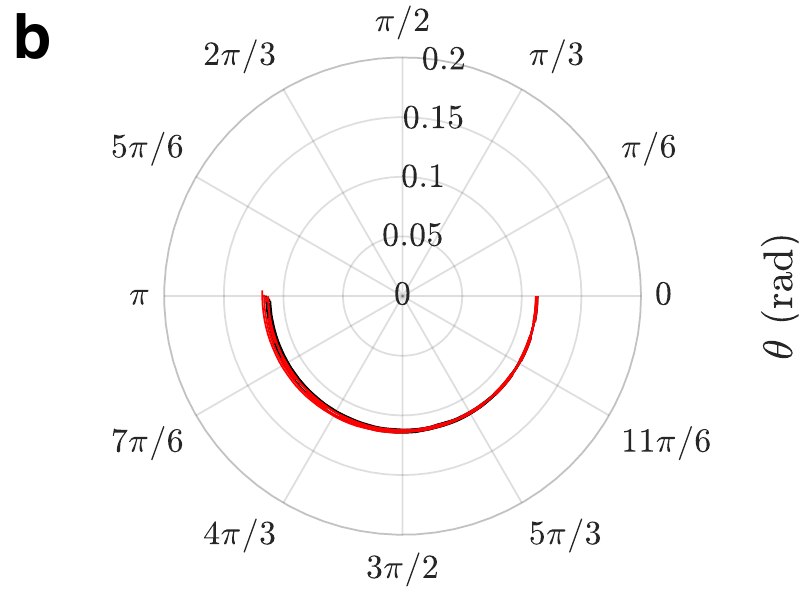}
\includegraphics[width=1.6in]{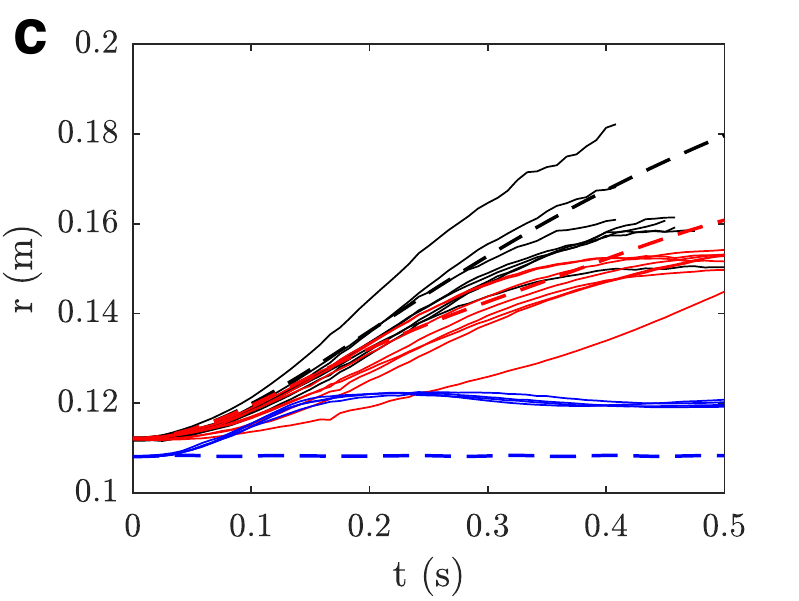}
\includegraphics[width=1.6in]{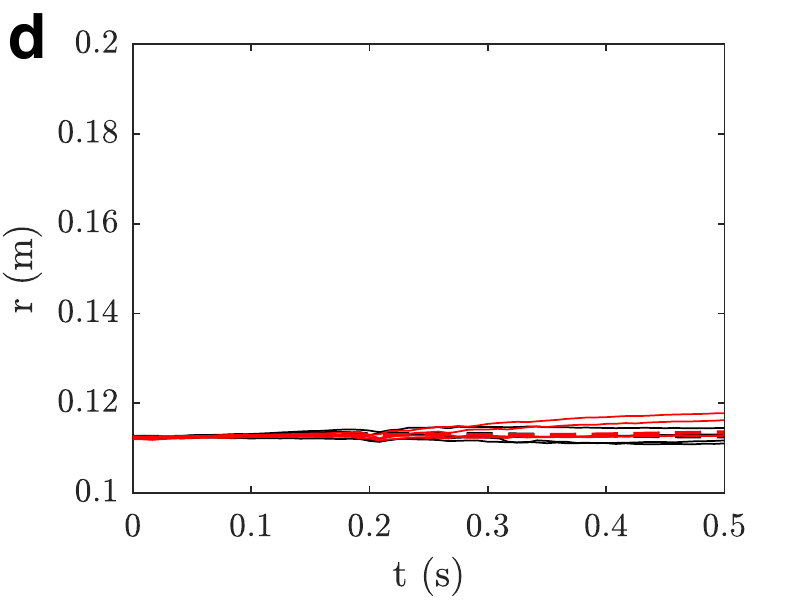}
\includegraphics[width=1.6in]{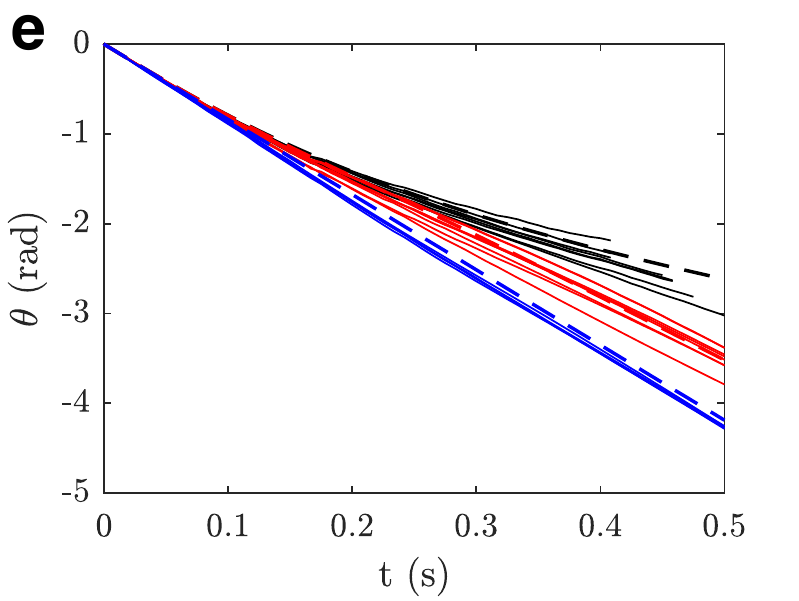}
\includegraphics[width=1.6in]{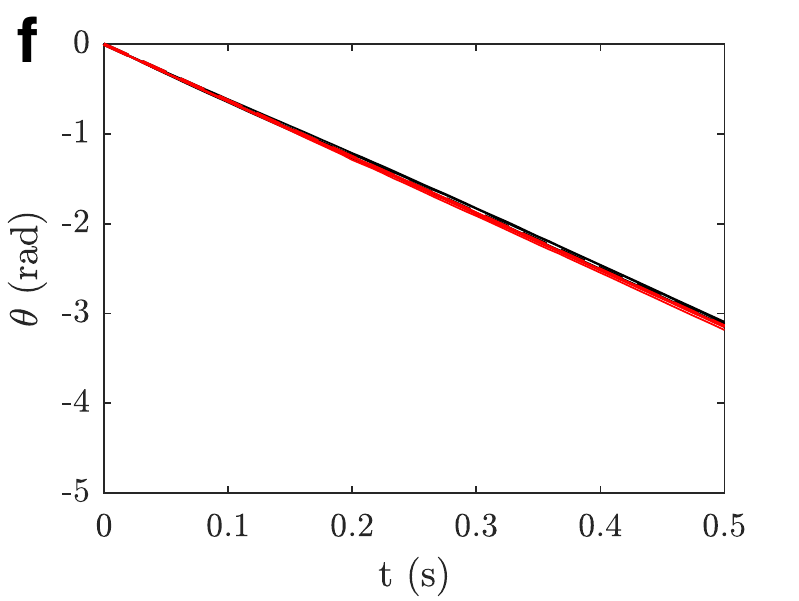}

\caption{Ball trajectories in polar coordinate and their time evolution in the lab frame for two different rotation profiles.
\textbf{a} Ball trajectories in the quasi-Keplerian (qK) flows $(\mathit{\Omega}_1, \mathit{\Omega}_3, \mathit{\Omega}_2) = (190, 80, 22)$~rpm clockwise, for the angular speeds of inner cylinder, local post frame, and outer cylinder, respectively.
\textbf{b} Ball trajectories in the solid body flows $(\mathit{\Omega}_1, \mathit{\Omega}_3, \mathit{\Omega}_2) =(60, 60, 60)$~rpm clockwise. \textbf{c} (\textbf{d}) Time evolution of radial coordinates for the qK (solid body) flows. \textbf{e} (\textbf{f}) Time evolution of azimuthal coordinates for the qK (solid body) flows. Experimental results for untethered, weak spring-tethered and, strong spring tethered, cases are shown in black, red, and blue respectively. Predictions from solving Eqs.~(\ref{hill1}) and (\ref{hill2}) for each of these cases are shown as corresponding dashed lines. }
\label{Trajectory}
\end{figure}

\begin{figure}[htb]
\includegraphics[width=1.6in]{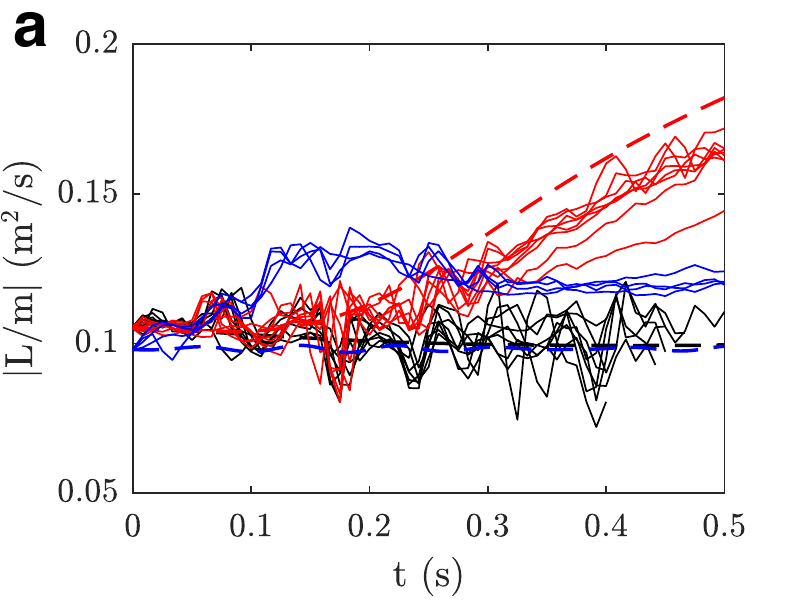}
\includegraphics[width=1.6in]{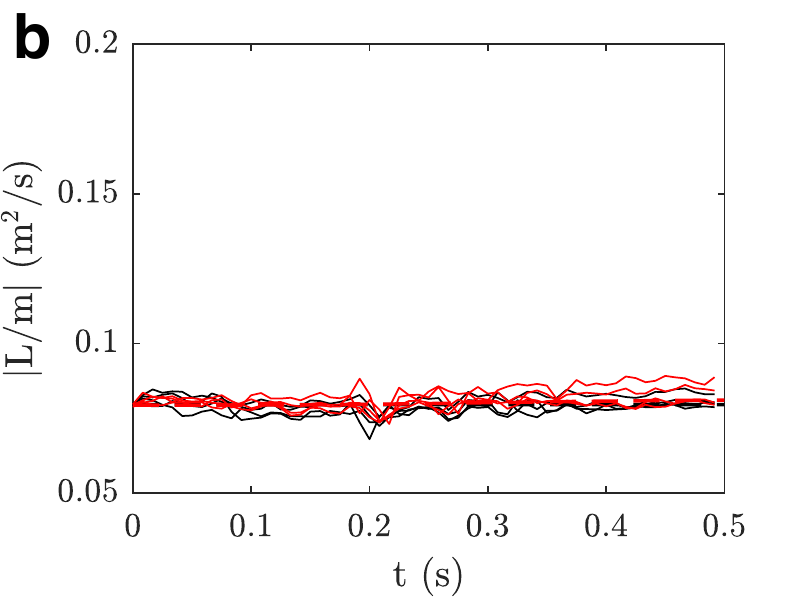}
\includegraphics[width=1.6in]{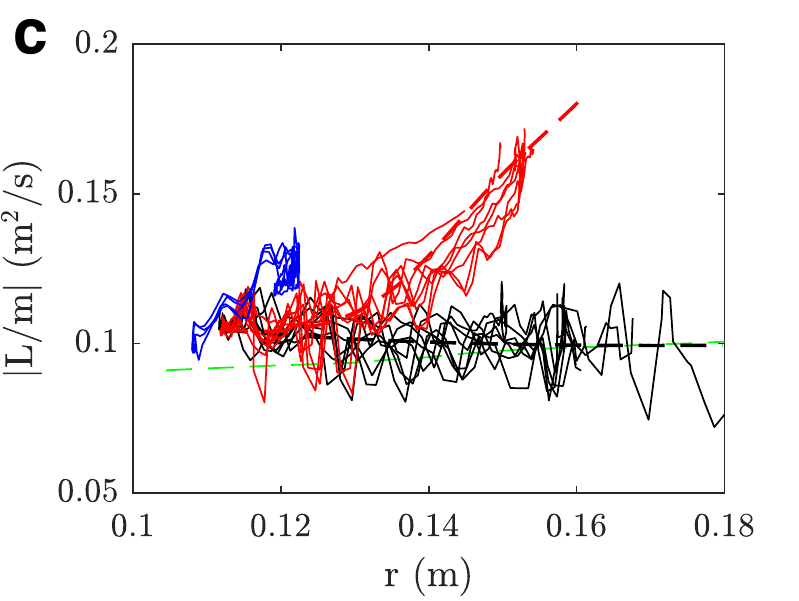}
\includegraphics[width=1.6in]{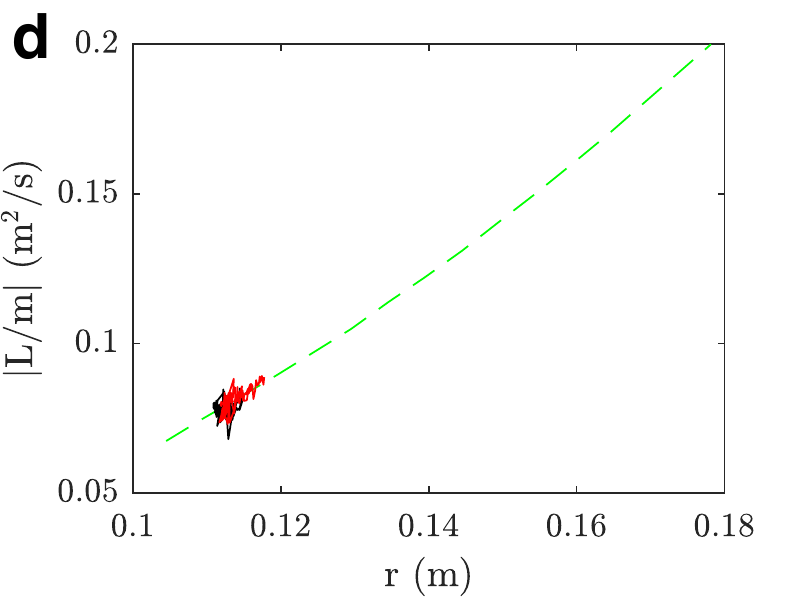}
\caption{Time evolution and radial evolution of the ball's angular momentum.
\textbf{a} (\textbf{b}) Time evolution of the ball's angular momentum in the quasi-Keplerian or qK (solid body) flows.
\textbf{c} (\textbf{d}) Radial evolution of the ball's angular momentum in the qK (solid body) flows.
Experimental results for untethered, weak spring-tethered and, strong spring tethered, cases are shown in black, red, and blue respectively. Predictions from solving Eqs.~(\ref{hill1}) and (\ref{hill2}) for each of these cases are shown as corresponding dashed lines. Efficient angular momentum transport occurs only in the case of the qK flows using a weak-spring tether (red). All other cases show little angular momentum transport, as expected. The green dashed line shows the background flow angular momentum profile.}
 \label{AngularMomentum}
\end{figure}

\bigskip
\noindent
\textbf{Experimental measurements.} 
For solid-body ($q=0$) and qK ($0<q<2$) flows, we compare the motion of an untethered ball to that of a ball tethered to a post anchored at a local rotating frame ($\mathit{\Omega}_3$= 80 rpm, clockwise) by a weak or strong spring. These cases are listed in Table~\ref{growth}.

Figure \ref{Trajectory} shows polar coordinate and time dependent ball trajectories in the lab frame. Each solid line of a given color corresponds to a separate experimental run with the same initial conditions. The left and right column panels correspond to qK and solid-body flow cases respectively. For each run in the qK case, the ball is initially held to the post rotating at $\mathit{\Omega}_3$ which rotates slightly faster (and has more angular momentum) than the background flow at its radius, to minimize secondary Ekman flow, as in the cases with both caps \cite{Edlund14}. The ball therefore drifts to larger radii, regardless of whether it is tethered or untethered. However, the ball lags behind less in azimuth in the rotating frame for the tethered cases and thus advances ahead to more negative angles in the lab frame (Fig.~\ref{Trajectory}a). The radial and azimuthal drift speeds are also different for tethered versus untethered cases. The radial velocity is lower for the tethered than untethered cases (Fig.~\ref{Trajectory}c). The tethered cases exhibit faster angular speeds, as evidenced by their steeper slopes in Fig.~\ref{Trajectory}e. 

The dashed lines show the corresponding solutions to Eqs.~(\ref{hill1}) and (\ref{hill2}). 
Amplitudes of oscillation modes across all presented cases are negligible compared to experimental noise. The very early time linear growth rate, within the noise, is consistent with the standard MRI growth rate with negligible Stokes drag $D_1$. At late times, saturation from nonlinear damping by the $D_2$ term is most consistent with the data. 

Most telling are the specific angular momentum evolution plots of Fig.~\ref{AngularMomentum}. Figure~\ref{AngularMomentum}a shows that for the qK flows, the angular momentum of the ball remains constant for the untethered case (solid black lines) as expected from angular momentum conservation. In contrast, the weak spring tethered ball gains angular momentum (solid red lines) as expected from the MRI. Fig.~\ref{AngularMomentum}c correspondingly shows that the tethered ball gains angular momentum as it moves outward.

For solid body flow, Fig.~\ref{Trajectory}d shows that the ball hardly moves in radius from its initial position for either the weak spring case (red) or the untethered case (black). Correspondingly, Fig.~\ref{AngularMomentum}b and Fig.~\ref{AngularMomentum}d show little difference in the red and black lines for  solid-body flow runs. The blue lines in the plots of Figs.~\ref{Trajectory} and \ref{AngularMomentum} show the case of a strong spring where the MRI mechanism is predicted to be ineffective. All of these blue trajectories are consistent with theoretical expectation that outward motion is halted once the strong spring is taut and angular momentum transfer is abated. The initial radial drift and associated angular momentum gain in the strong spring case is due to a limitation of the experimental setup, namely that the spring anchor point is offset from the center of mass of the ball. This does not affect the physics conclusions.

\bigskip
\noindent
\textbf{\large Discussion} 

\noindent
While many astrophysical processes are difficult to test and validate in the lab, theory should be experimentally validated when possible and this is one of the core pillars of the discipline of laboratory astrophysics. In this context, neither the standard MRI instability, nor its mechanical analogue have been previously demonstrated in the laboratory, despite their widespread use in theoretical astrophysics.
Measurements from our new apparatus now experimentally confirm the mechanism of angular momentum transport by the MRI and thus support its validity.

The measurements are all consistent with the theoretical implications of Eqs.~(\ref{hill1}) and (\ref{hill2}). Specifically, (i) only for the weak spring case with a qK $(0<q<2)$ flow, does the MRI-like instability manifest, and sustain angular momentum transport from post to ball; 
(ii) measured trajectories of the ball agree with non-linear model equations for weak-spring tethered, strong-spring tethered, and untethered cases for qK and solid-body flows; (iii) Reynolds drag eventually balances the spring force to saturate the instability in the tethered case.  
Larger experiments could better distinguish linear from non-linear regimes and detailed investigations could further delineate the ``weak'' and ``strong'' spring  transition. 

Our spring-ball apparatus highlights use of a novel combination of solid and fluid mechanics 
to test MHD principles in the lab. The apparatus requires careful choices of the experimental parameters to ensure that the MHD analogue is captured: the dominant forces governing the motion of the ball must directly correspond to the dominant forces governing the motion of a parcel of MHD fluid for the chosen experimental design.

\begin{figure}[b]
\raisebox{-0.5\height}{\includegraphics[width=1.8in]{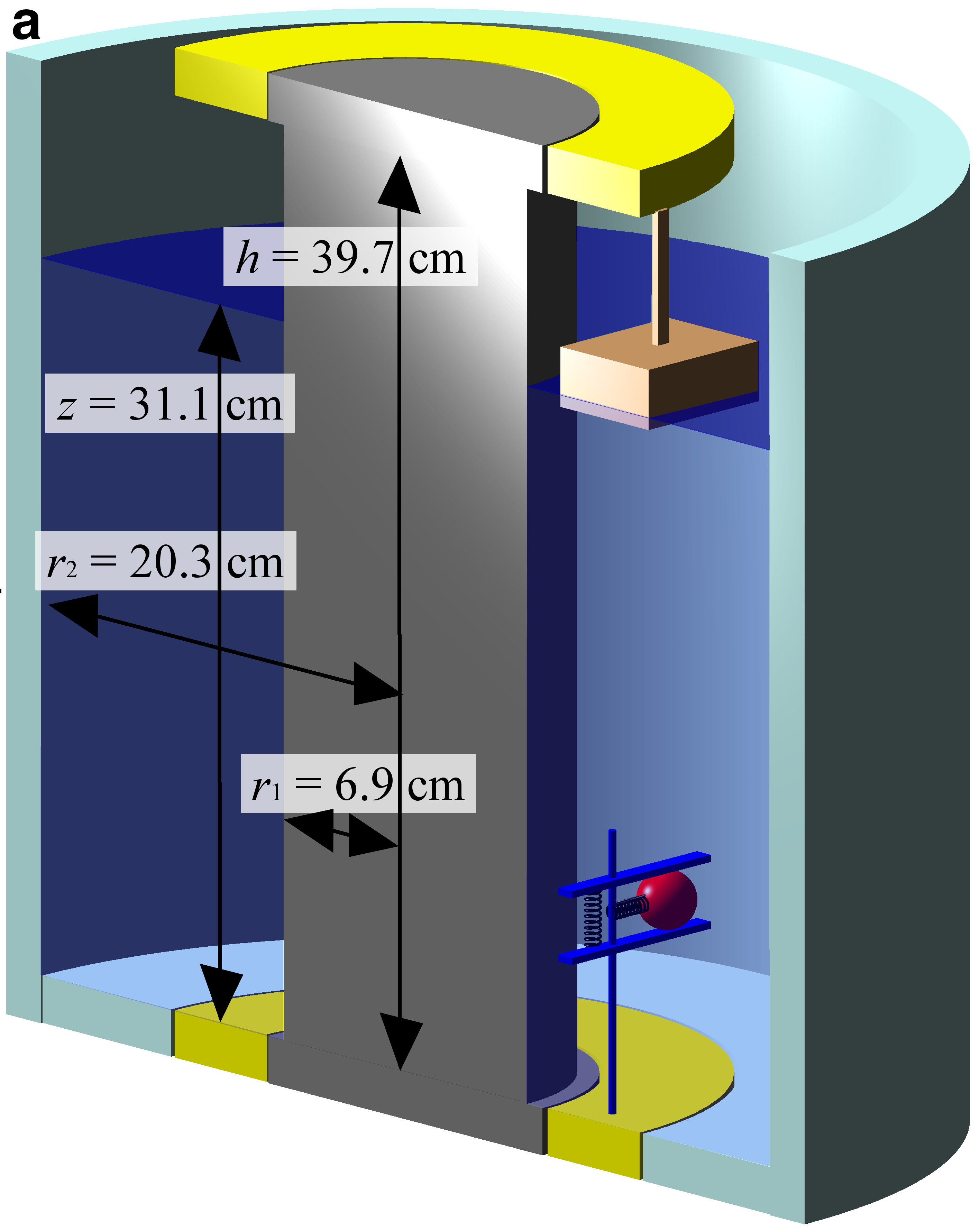}}
\raisebox{-0.5\height}{\includegraphics[width=0.81in]{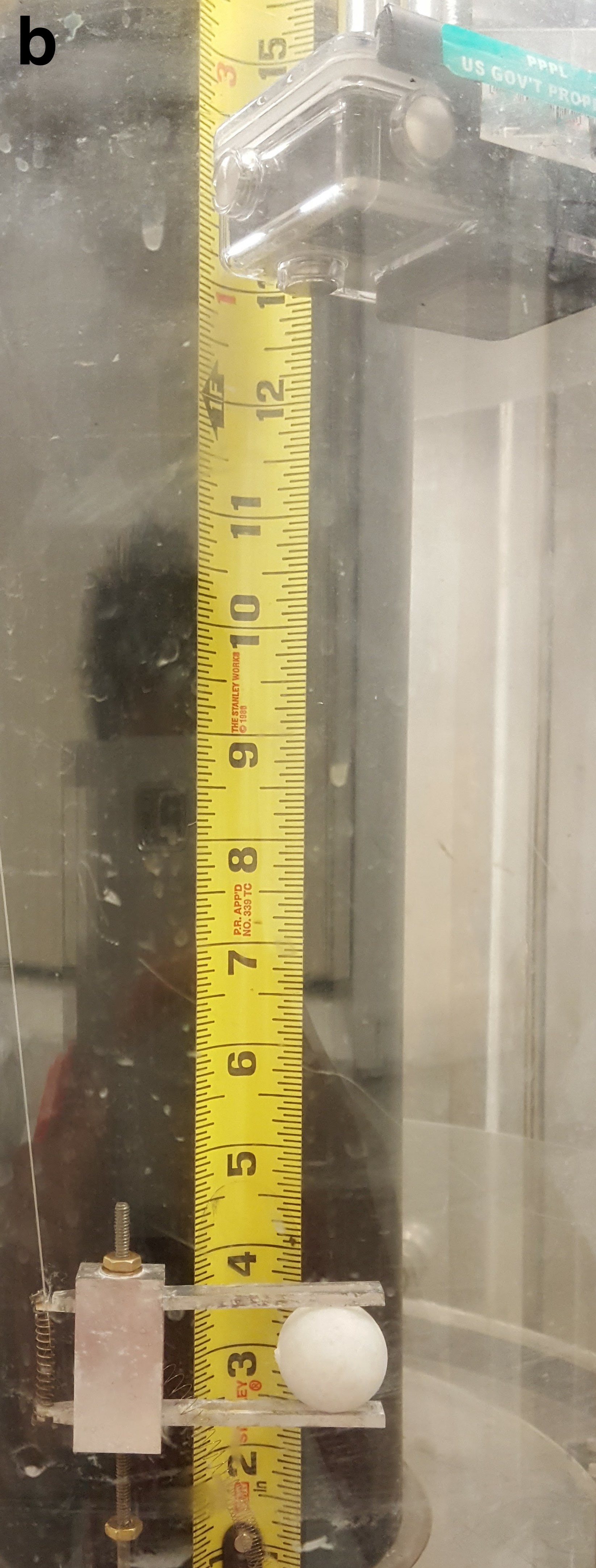}}
\caption{Experimental apparatus and diagnostics. \textbf{a} Schematic of the modified Taylor-Couette device~\cite{Edlund14}, with the inner and outer cylinder radii of $r_1 = 6.9$~cm and $r_2 = 20.3$~cm respectively, and the height of $h = 39.7$~cm. The device was filled with water to a depth of 31.1 cm and the top was open to allow access. A GoPro HERO4 camera was partially submerged to minimize optical distortion.
The camera was supported by an attachment and co-rotated with the ring (yellow) at $\mathit{\Omega}_3$. A 1-inch-diameter neutrally buoyant test mass (red) was tethered by an unstretched spring to a vertical post and held by a spring-loaded jaw-clamp. At $t=0$ a release line (not shown) was pulled from above allowing the vertical spring to relax, releasing the mass. \textbf{b} Photograph of the test mass and release mechanism. The camera is visible at the top of the image.}
\label{Schematic}
\end{figure}

\bigskip
\noindent
\textbf{\large Methods}

\noindent
\textbf{Apparatus.}
The experiments were carried out in a modified Taylor-Couette device (Fig.~\ref{Schematic}) using water and an open top cap. Two co-axial cylinders with height $h = 39.7$~cm, and radii $r_1 = 6.9$~cm and $r_2 = 20.3$~cm, were driven by motors at two independent angular rotation rates $\mathit{\Omega}_1$ and $\mathit{\Omega}_2$. qK flows in which $\mathit{\Omega}_1 > \mathit{\Omega}_2$ while $\mathit{\Omega}_1 r_1^2 < \mathit{\Omega}_2 r_2^2$ can be established. To minimize secondary Ekman flow, 
axial boundaries are divided into three annuli. The innermost annulus with $r < 8$~cm co-rotates with the inner cylinder while the outermost annulus with $r > 14$~cm co-rotates with the outer cylinder. The intermediate annulus where $8\textrm{ cm} < r < 14\textrm{ cm}$ is driven by a third motor at a rotation rate $\mathit{\Omega}_3$. The secondary flow can be minimized by a suitable choice of $\mathit{\Omega}_3$, resulting in an extremely quiescent qK flow \cite{Edlund14}. Our experiments used only the bottom boundary, allowing top access to the interior. To avoid significant fluid height variation that occurs on a rotating free surface, the rotation rates were limited to $\mathit{\Omega}_1 = 190$~rpm, $\mathit{\Omega}_3 = 80$~rpm, and $\mathit{\Omega}_2 = 22$~rpm. Measurements of the azimuthal velocity at the mid-height of the fluid using laser Doppler velocimetry confirmed that the flow had nearly the ideal Couette profile with negligible Ekman effect (as using both axial boundaries \cite{Edlund14}) with $q \leq2$ with little dependence on $r$ and $z$. Practical limitations on rotation rates and spring constants led us to use 1-inch diameter water-filled plastic spheres, of total mass 8.43~g. With any tethering spring, they were nearly neutrally buoyant. The finite size of the spherical test masses, as compared with $r_1$ and $r_2$ is included in the analysis 
as discussed above. The test mass was held in place by a clamp attached to a vertical post mounted at $r_0 = 10.8$~cm on the annular ring rotating at $\mathit{\Omega}_3$. This radius was originally selected so that $\mathit{\Omega}_3 = \mathit{\Omega}_\mathrm{TC}(r_0)$ where $\mathit{\Omega}_\mathrm{TC}(r)$ is the ideal Couette profile with a $\mathit{\Omega}_1:\mathit{\Omega}_3:\mathit{\Omega}_2 = 190:80:22$. The height $l = 12.7$~cm of the vertical post was chosen so that the test mass would sit away from the lower boundary and the top surface at $z = 31.1$~cm. The clamp release was triggered by hand using a metal arm fixed in the laboratory frame.
The test mass was either untethered to the vertical post, or tethered with either a weak or strong spring. The springs had measured spring constants of $k_\mathrm{weak} = 0.636$~N m$^{-1}$ and $k_\mathrm{strong} = 51.5$~N m$^{-1}$. We estimate the effective Reynolds number of the flow around the ball using $Re = 2 R [ \dot r ^2 + r ^2 [\dot {\theta} - \mathit{\Omega}(r)] ^2 ] ^{1/2} / \nu_{H_2O}$, and find maximum values $Re \approx 5000 - 20000$ for qK runs and $Re \approx 1$ for solid body. The former values are consistent with the importance of the $D_2$ term in Eqs.~(\ref{hill1}) and (\ref{hill2}).

\bigskip
\noindent
\textbf{Diagnostics.}
We mounted a compact battery-powered, waterproof, video camera in the rotating frame of the vertical post with rotation rate $\mathit{\Omega}_3$ so that the test mass appeared stationary until release at $t = 0$. The camera captured 120 frames per second and the lens was slightly immersed in the water to minimize further optical distortions due to the fluid free surface. After each run, the recorded video was transferred to a computer. The camera uses a ``fisheye'' lens for a wide field-of-view, but this distortion was readily removed using commonly available software. The location of the center of the test mass in each frame was determined automatically by object identification and tracking software.  Cartesian image data were converted into polar coordinates. From the position data, velocities, acceleration, and the vertical component of the angular momentum were calculated. The accuracy of the position data is limited by factors such as motion blur, tracking errors, the abilities to correct for lens distortion and refraction.

\bigskip
\noindent
\textbf{\large Data availability}

\noindent
The digital data for this paper can be found at \url{http://arks.princeton.edu/ark:/88435/dsp01x920g025r}.

\bibliography{MRI-ball.bib}
\bibliographystyle{naturemag}

\bigskip
\noindent
\textbf{\large Acknowledgments}

\noindent
D.H., K.C., E.G. and H.J. acknowledge support from NASA (NNH15AB25I), NSF (AST-1312463) and DoE (DE-AC0209CH11466). E.B. acknowledges support from the Simons Foundation and the Institute for Advanced Study (Princeton) while on sabbatical, and grants NSF-AST-15156489 and HST-AR-13916, the Kavli Institute for Theoretical Physics (KITP) USCB with associated support from grant NSF PHY-1125915. Authors acknowledge technical support by Peter Sloboda.

\bigskip
\noindent
\textbf{\large Author contributions}

\noindent
E.B. and H.J. initiated the research.
D.H. modified the apparatus and conducted the experiments with the help of K.C. and E.G., guided by H.J. and E.B. 
D.H. analyzed the data, performed theoretical calculations and generated result figures with the guidance of all other authors.
H.J. generated analogue diagrams and E.G. generated apparatus figure.
E.B., D.H., E.G. and H.J. drafted and revised the manuscript. All authors discussed the results and interpretations.

\bigskip
\noindent
\textbf{\large Additional information}

\noindent
\textbf{Competing interests:} The authors declare no competing interests.

\end{document}